\documentclass[preprintnumbers,amsmath,amssymb,floatfix,12pt,superscriptaddress,nofootinbib]{revtex4}

\begin{document}
\title{Interacting entropy-corrected agegraphic Chaplygin gas model of dark energy}

\author{M. Malekjani\footnote{Email:\text{malekjani@basu.ac.ir}}}
\author{A.Khodam-Mohammadi\footnote{Email:\text{khodam@basu.ac.ir}}}
\affiliation{Department of Physics, Faculty of Science, Bu-Ali Sina
University, Hamadan 65178, Iran}

\begin{abstract}
\vspace*{1.5cm} \centerline{\bf Abstract} \vspace*{1cm} In this
work, we consider the interacting agegraphic dark energy models with
entropy correction terms due to loop quantum gravity. We study the
correspondence between the Chaplygin gas energy density with the
interacting entropy-corrected  agegraphic dark energy models in
non-flat FRW universe. We reconstruct the potentials and the
dynamics of the interacting entropy-corrected agegraphic scalar
field models. This model is also extended to the interacting
entropy-corrected agegraphic generalized Chaplygin gas dark energy.
\end{abstract}

\maketitle

\newpage
\section{Introduction}
 The dark energy(DE) problem is one of the most famous problems in modern cosmology
since the discovery of accelerated expansion of the universe by supernova cosmology project and
high redshift supernova experiments \cite{rei}
. The WMAP (Wilkinson Microwave Anisotropy Probe) experiment \cite{wmap},
indicates that the dark energy occupies about 70 \% of the total
energy of the universe. The
simplest candidate for dark energy is the cosmological constant in
which the equation of state is independent of the cosmic time.
However, two problems arise from the cosmological constant, namely
the fine-tuning and the cosmic coincidence problems \cite{copel}. In
order to alleviate or even solve these problems, many dynamical
dark energy models have been suggested, whose equation of state
evolves with cosmic time.  i) The dynamical dark energy can be
realized by scalar fields. Scalar field models arise in string
theory and are studied as a candidates for dark energy. It includes
quintessence \cite{Wetterich}, K-essence \cite{Chiba}, phantoms
\cite{Caldwell1}, tachyon \cite{Sen}, dilaton \cite{Gasperini},
quintom \cite{Elizalde1} and so forth. ii) The interacting dark
energy models, by considering the interaction between dark matter and
dark energy, including Chaplygin gas \cite{Kamenshchik}, braneworld
models \cite{Deffayet}, holographic DE  and agegraphic DE models ,
etc.\\
The holographic dark energy model (HDE) is one of
interesting candidates for dark energy, which have been suggested
based on the holographic principle. This model has been extensively
investigated in the literature \cite{coh99,hora,li04}.
 In the principle of holographic,
the energy density depends on the entropy-area relationship of black
holes in Einstein gravity \cite{8,hora}. This entropy is given as
$S_{\rm BH} = A/(4G)$, where $A\sim L^2$ is the area of horizon.
The energy density of HDE is given by
\begin{equation}
\rho_D=3n^2M_p^2L^{-2},
\end{equation}
where $n$ is a numerical factor, $M_p$ is the reduced plank mass and $L$ is a length scale.
If we take the length scale $L$ as a size of current universe or
particle horizon, the accelerated expansion of the universe can not
be derived by HDE model\cite{hsu}. However, in the case of event
horizon as a length scale, HDE model can derive the universe with
accelerated expansion \cite{li04}. The problem arises from the event
horizon length scale is that the it is a global concept
of spacetime and existence of it depends on the future evolution of
the universe only for universe with forever accelerated universe.
Furthermore, the holographic with event horizon as a length scale is
not compatible with the age of some old high redshift objects
\cite{zhang4}. To avoid the problem of causality, appearing with
event horizon area as a IR cut-off, Garanda and Oliviors proposed a
new IR cut-off for the HDE containing the Hubble and time derivative
Hubble scales \cite{garanda}. This new model of HDE depends on local
quantities.\\

In the context of loop quantum gravity (LQG), the
entropy-area relationship, $S_{\rm BH} = A/(4G)$,  can be modified from the inclusion of
quantum effects. The quantum corrections provided to the
entropy-area relationship leads to the curvature correction in the
Einstein-Hilbert action and vice versa \cite{zhu}. The corrected
entropy is given by \cite{modak}
\begin{equation}
S_{\rm
BH}=\frac{A}{4G}+\tilde{\alpha}\ln{\frac{A}{4G}}+\tilde{\beta},\label{MEAR}
\end{equation}
where $\tilde{\alpha}$ and $\tilde{\beta}$ are dimensionless
constants of order unity. Considering the entropy correction, the
energy density of entropy-corrected holographic dark energy (ECHDE)
can be given as \cite{wei1}
\begin{equation}\label{rhoS}
\rho _{D }=3n^2M_{p}^{2}L^{-2}+\alpha L^{-4}\ln
(M_{p}^{2}L^{2})+\beta L^{-4},
\end{equation}
where $\alpha$ and $\beta$ are the dimensionless constant. The second and third terms are due
to entropy correction. Putting $\alpha=\beta=0$, the energy density of ECHDE reduces to the energy density of
ordinary HDE.\\
Recently, based on principle of quantum gravity, the
 agegraphic dark energy (ADE) and the new agegraphic dark
energy (NADE) models were proposed by Cai \cite{Cai1} and Wei \& Cai
\cite{cai4}, respectively. The ADE model is based on the line of
quantum fluctuations of spacetime, the so-called K\'{a}rolyh\'{a}zy
relation $\delta t=\lambda t_{p}^{2/3}t^{1/3}$, and the energy-time
Heisenberg uncertainty relation $E_{\delta t^{3}}\sim t^{-1}$. These
relations enable one to obtain an energy density of the metric
quantum fluctuations of Minkowski spacetime as follows \cite{Maz}
\begin{equation}
\rho _{q}\sim \frac{E_{\delta t^{3}}}{\delta t^{3}}\sim \frac{1}{%
t_{p}^{2}t^{2}}\sim \frac{M_{p}^{2}}{t^{2}}.  \label{ED}
\end{equation}%
where $t_p$ and $M_p$ are the reduced plank time and plank mass, respectively.
Choosing the age of universe, as a length measure, the casuality problem in HDE is avoided \cite{Cai1}.
In ADE model the energy density of dark energy is given by Eq.(\ref{ED}).
 However, in Friedmann-Robertson-Walker (FRW) universe, due to effect of
curvature, one should assume a numerical factor $3n^{2}$ in
Eq.(\ref{ED}) \cite{Cai1}. The new model of agegraphic dark energy
(NADE) has been proposed by Wei and Cai \cite{cai4}, in which the
cosmic time is replaced by the conformal time. The ADE and NADE models
have been constrained by various astronomical observations \cite{age,shey1,shey2,setare,age2}.\\
Regarding the entropy correction due to loop quantum gravity,
the agegraphic dark energy models have been
investigated . Taking the entropy
correction into account, Jamil \& Sheykhi generalized the agegraphic
tachyon models of dark energy \cite{jamil}. Karami \& sorouri
extended the agegraphic dark energy into the entropy-corrected
agegraphic dark energy in non flat universe \cite{karam}. Faroog, et
al.\cite{faroog} studied the correspondence between the tachyon,
K-essence and dilaton scalar field models with the interacting
entropy-corrected new agegraphic dark energy model in
the non-flat FRW universe.\\
Here, in this work we consider the entropy corrected agegraphic dark energy model.
 The energy density of entropy-corrected
agegraphic dark energy (ECADE) can be easily obtained by replacing
$L$ in Eq.(\ref{rhoS}) with a time scale $T$ of the universe.
Hence, the energy density of ECADE is given as
\begin{equation}\label{rhoS1}
\rho_D=3n^2M_p^2T^{-2}+\alpha T^{-4}\ln(M_p^2T^2)+\beta T^{-4}.
\end{equation}
The last terms in Eq.(\ref{rhoS1}) can be comparable to the first term at the early epoch
 of the universe and negligible when the time scale $T$ becomes large. Therefore, the entropy
correction can be important at the early time and the ECADE model reduces to the ordinary ADE model, when $T$ becomes large.\\
 On the other
hand, the Chaplygin gas model is one of the candidate of dark energy
models to explain the accelerated expansion of the universe. The
Chaplygin gas dark energy model can be assumed as a possible unification of
dark matter and dark energy. The equation of state of a prefect
fluid, Chaplygin gas, is given by
\begin{eqnarray}\label{Chap}
 p_D^{ch}=\frac{-A}{\rho_D^{ch}},
 \end{eqnarray}
 where A is a positive constant, $P_d^{ch}$ and $\rho_d^{ch}$ are the pressure and the energy
density of the Chaplygin gas dark energy, respectively. Chaplygin
gas plays a dual role at different epoch of the history of the
universe: it can be as a dustlike matter in the early time (i.e. for
small scale factor a), and as a cosmological constant at late times
(i.e. for large values of a). This model from the field theory
points of view are investigated in \cite{Bil}. The Chaplygin gas
emerges as an effective fluid associated with D-branes \cite{Bor}
and can also be
obtained from the Born-Infeld action \cite{Ben}.\\
The aim of this paper is to establish a correspondence between the
interacting ECADE and ECNADE ( entropy-corrected new agegraphic dark
energy) scenarios with the Chaplygin gas model in a non-flat
universe. The non-flatness of the universe is favored by astronomical
 observations \cite{seljak,wmap}. We propose the entropy-corrected agegraphic descriptions
of the Chaplygin gas dark energy and reconstruct the potential and
the dynamics of the scalar field which describe the Chaplygin
cosmology. This work is also extended to the generalized Chaplyin
gas model.

\section{THE INTERACTING ECADE as a Chaplygin gas \label{ORI}}
The Friedman-Robertson-Walker (FRW) metric for a universe with
curvature $k$ is given by
\begin{eqnarray}
 ds^2=dt^2-a^2(t)\left(\frac{dr^2}{1-kr^2}+r^2d\Omega^2\right),\label{metric}
 \end{eqnarray}
where $a(t)$ is the scale factor, and $k = -1, 0, 1$ represents the
open, flat, and closed universes, respectively.  The freidmann
equation for a non-flat universe containing dark energy (DE) and
cold dark matter (CDM) is written as
\begin{eqnarray}\label{Fried}
H^2+\frac{k}{a^2}=\frac{1}{3M_p^2} \left( \rho_m+\rho_D \right).
\end{eqnarray}

Let us define the dimensionless energy densities as
\begin{eqnarray}\label{Omega}
\Omega_m=\frac{\rho_m}{\rho_c}, \hspace{0.5cm}
\Omega_D=\frac{\rho_D}{\rho_c},\hspace{0.5cm} \Omega_k=\frac{k}{H^2
a^2},
\end{eqnarray}
where $\rho_c=3M_p^2H^2$ is a critical energy density. Thus, the
Friedmann equation with respect to the fractional energy densities
can be written as
\begin{eqnarray}\label{Fried2}
\Omega_m+\Omega_D=1+\Omega_k.
\end{eqnarray}
substituting the equation of state of Chaplygin gas (i.e., Eq.
\ref{Chap}) into the relativistic energy conservation equation,
leads to a evolving density as
\begin{eqnarray}\label{rhochap}
\rho_D^{ch}=\sqrt{A+\frac{B}{a^6}}.
\end{eqnarray}
where $B$ is an integration constant. The energy density and
pressure of the scalar field, regarding the Chaplygin gas dark
energy is written as
\begin{eqnarray}\label{rhophi}
\rho_\phi=\frac{1}{2}\dot{\phi}^2+V(\phi)=\sqrt{A+\frac{B}{a^6}},\\
p_\phi=\frac{1}{2}\dot{\phi}^2-V(\phi)=\frac{-A}{\sqrt{A+\frac{B}{a^6}}},
\label{pphi}
\end{eqnarray}
Hence, it is easy to obtained the scalar potential and the kinetic
energy terms for the Chaplygin gas as
\begin{eqnarray}\label{vphi}
&&V(\phi)=\frac{2Aa^6+B}{2a^6\sqrt{A+\frac{B}{a^6}}},\\
&&\dot{\phi}^2=\frac{B}{a^6\sqrt{A+\frac{B}{a^6}}}. \label{ddotphi}
\end{eqnarray}

Now we reconstruct the ECADE Chaplygin gas dark energy model. In
order to do this, first we describe the ECADE model and then
reconstruct the ECADE Chaplygin gas model. The energy density of the
ECADE is given by Eq.(\ref{rhoS1}) where the time scale $t$ is
chosen as the age of the universe,$T$. Therefore, the energy density
of ECADE is written as
\begin{equation}\label{rhoori}
\rho_D=3n^2M_p^2T^{-2}+\alpha T^{-4}\ln(M_p^2T^2)+\beta T^{-4},
\end{equation}
where $T$ is defined by
\begin{equation}
T=\int\limits_0^a\frac{da}{aH}.
\end{equation}

Using the Eqs.(\ref{Omega}, \ref{rhoori}), it is easy to find that
\begin{equation}
\Omega_D=\frac{n^2}{H^2T^2}+\frac{\alpha}{3M_p^2H^2T^4}\ln(M_p^2T^2)
+\frac{\beta}{3M_p^2H^2T^4}.
\end{equation}
Differentiating Eq.(\ref{rhoori}) with respect to cosmic time $t$,
obtains
\begin{equation}\label{dotrho}
\dot{\rho}_D=-2H\left(\frac{3n^2M_p^2T^{-2}+2\alpha
T^{-4}\ln(M_p^2T^2)+(2\beta-\alpha) T^{-4}}{\sqrt{3n^2M_p^2+\alpha
T^{-2}\ln(M_p^2T^2)+\beta T^{-2}}}\right)\sqrt{3M_p^2\Omega_D}.
\end{equation}
Considering the interaction between dark matter and dark energy, the
energy conservation equations for ECADE and CDM are
\begin{eqnarray}
\dot\rho_m+3H\rho_m&=&Q,\\
\dot\rho_D+3H\rho_D(1+w_D)&=&-Q,\label{energy_conserv}
\end{eqnarray}
where $Q=3b^2H\rho$ denotes the interaction between dark matter
and dark energy and $b^2$ is a coupling parameter. Inserting
Eq.(\ref{dotrho}) in (\ref{energy_conserv}) yields the EoS
parameter of interacting ECADE as
\begin{equation}\label{w_d_ECADE}
w_D=-1+\frac{2}{3}\frac{T(2D_T-3n^2M_p^2T^2-\alpha)\sqrt{3M_p^2\Omega_D}}{D_T^{3/2}}-\frac{b^2(1+\Omega_k)}{\Omega_D}.
\end{equation}
where $D_T=3n^2M_p^2T^2+\alpha
\ln({M_p^2T^2})+\beta=\rho_DT^4=\Omega_D\rho_cT^4$. It should be
noted that the parameter $\beta$ is covered in definition of
$D_T$. In the limiting case of $\alpha=\beta=0$, representing
the interacting ADE model without the entropy correction,
$D_T=3n^2M_p^2T^2$ and the Eq.(\ref{w_d_ECADE}) reduces into
the simple form as
\begin{equation}\label{ade_ordinary}
w_D=-1+\frac{2}{3}\sqrt{\frac{\Omega_D}{n^2}}-\frac{b^2(1+\Omega_k)}{\Omega_D}.
\end{equation}
Here, Eq.(\ref{ade_ordinary}) is same as Eq.(14) in
Ref.\cite{shayekh}, for the EoS parameter of ordinary agegrpahic
dark energy model.
 Now, we can construct the interacting ECADE Chpalygin
gas model. By equating the Eqs.(\ref{rhochap}) and (\ref{rhoori}),
we have
\begin{equation}\label{B}
B=\frac{a^6(-AT^8+D_T^2)}{T^8},
\end{equation}
 Using
Eqs.(\ref{Chap}), (\ref{rhochap}) and (\ref{w_d_ECADE}) one can
obtain
\begin{eqnarray}\label{khafan}
w_D&=&\frac{-A}{\rho_D^{ch}}=\frac{-A}{A+Ba^{-6}}\\ \nonumber &&
=-1+\frac{2}{3}\frac{T(2D_T-3n^2M_p^2T^2-\alpha)\sqrt{3M_p^2\Omega_D}}{D_T^{3/2}}-\frac{b^2(1+\Omega_k)}{\Omega_D}.
\end{eqnarray}
Substituting $B$ in Eq.(\ref{khafan}), we obtain the constant $A$ as
\begin{eqnarray}
A=\frac{1}{3T^8}\label{A}
\left[2T\sqrt{3M_p^2\Omega_DD_T}(-2D_T+3n^2M_p^2T^2+\alpha)+3D_T^2(1+b^2\frac{1+\Omega_k}{\Omega_D})\right].
\end{eqnarray}
Putting $\alpha=\beta=0$ and $b=0$, the relations (\ref{B}) and
(\ref{A}) can be easily reduced into Eqs.(22) and (24) in
Ref.\cite{shyekh_age} for ordinary ADE Chaplygin gas without
interaction parameter.
 Substituting $A$ and $B$ in Eqs.(\ref{rhophi}) and (\ref{vphi}),
we can rewrite the scalar potential and kinetic energy terms as
\begin{equation}
V(\phi)=\frac{\sqrt{\frac{3M_p^2\Omega_D}{D_T}}}{3T^3}(-2D_T+3n^2M_p^2T^2+\alpha)+\frac{D_T}{2}(\frac{2+b^2\frac{1+\Omega_k}{\Omega_D}}{T^4}).
\end{equation}

\begin{equation}
\dot{\phi}=\sqrt{\frac{2}{3T^3}\sqrt{\frac{3M_p^2\Omega_D}{D_T}}(+2D_T-3n^2M_p^2T^2-\alpha)-\frac{b^2}{T^4}D_T\frac{1+\Omega_k}{\Omega_D}}.
\end{equation}
We can also see that, the Eqs.(26) and (27) of Ref.\cite{shyekh_age}
can be achieved for $\alpha=\beta=0$ and $b=0$. Using
$\dot{\phi}=\phi^{\prime}H$, we have
\begin{equation}\label{phi_prime}
\phi^{\prime}=\frac{1}{H}\sqrt{\frac{2}{3T^3}\sqrt{\frac{3M_p^2\Omega_D}{D_T}}(+2D_T-3n^2M_p^2T^2-\alpha)-\frac{b^2}{T^4}D_T\frac{1+\Omega_k}{\Omega_D}}.
\end{equation}
Integrating Eq.(\ref{phi_prime}), one can obtain the evolutionary
form of the phantom scalar field as
\begin{equation}
\phi(a)-\phi(a_0)=\int_{\ln{a_0}}^{\ln{a}}
\frac{1}{H}\sqrt{\frac{2}{3T^3}\sqrt{\frac{3M_p^2\Omega_D}{D_T}}(+2D_T-3n^2M_p^2T^2-\alpha)-\frac{b^2}{T^4}D_T\frac{1+\Omega_k}{\Omega_D}}d\ln
{a}
\end{equation}
where $a_0$ is a present value of scale factor. Here we established
the correspondence between the interacting ADE model with  Chaplygin
gas model and reconstruct the potential and the dynamics of
interacting ECADE Chaplygin gas. For convince, putting
$\alpha=\beta=0$ and $b=0$ results the relations (28) and (29) of
Ref.\cite{shyekh_age}.

\section{THE INTERACTING ECNADE as a Chaplygin gas \label{ORI}}
In this section we construct the interacting entropy-corrected new
agegraphic dark energy (ECNADE) model. The energy density of ECNADE
is given by
\begin{equation}\label{den_ecnade}
\rho_D=3n^2M_p^2\eta^{-2}+\alpha \eta^{-4}\ln(M_p^2\eta^2)+\beta
\eta^{-4}.
\end{equation}
where $\eta$ is a conformal time given by
\begin{equation}\label{ecnade}
\eta=\int{\frac{dt}{a}}=\int_0^a{\frac{da}{Ha^2}}.
\end{equation}
Using the fractional energy density in Eq.(\ref{Omega}), the
Eq.(\ref{ecnade}) can be rewritten as
\begin{equation}\label{ss2}
\Omega_D=\frac{n^2}{H^2\eta^2}+\frac{\alpha}{3M_p^2H^2\eta^4}\ln(M_p^2\eta^2)
+\frac{\beta}{3M_p^2H^2\eta^4}.
\end{equation}
Differentiating Eq.(\ref{den_ecnade}) with respect to cosmic time
$t$, one can obtain
\begin{equation}\label{ddd}
\dot{\rho}_D=-\frac{2H}{a}\left(\frac{3n^2M_p^2\eta^{-2}+2\alpha
\eta^{-4}\ln(M_p^2\eta^2)+(2\beta-\alpha)
\eta^{-4}}{\sqrt{3n^2M_p^2+\alpha \eta^{-2}\ln(M_p^2\eta^2)+\beta
\eta^{-2}}}\right)\sqrt{3M_p^2\Omega_D}.
\end{equation}
Substituting Eq.(\ref{ddd}) in Eq.(\ref{energy_conserv}), we can
obtain the Eos parameter of interacting ECNADE as
\begin{equation}\label{cc}
w_D=-1+\frac{2}{3}\frac{\eta(2D_{\eta}-3n^2M_p^2\eta^2-\alpha)\sqrt{3M_p^2\Omega_D}}{D_{\eta}^{3/2}}-\frac{b^2(1+\Omega_k)}{\Omega_D}.
\end{equation}
where $D_{\eta}=3n^2M_p^2\eta^2+\alpha
\ln({M_p^2\eta^2})+\beta=\rho_D\eta^4=\Omega_D\rho_c\eta^4$. Here
$\rho_D$ is the energy density of ECNADE. Now, we can construct the
interacting ECNADE model. Equating Eqs.(\ref{cc}) and
(\ref{rhochap}), the constant $B$ in Chaplygin gas model can be
obtained as
\begin{equation}\label{B2}
B=\frac{a^6(-A\eta^8+D_{\eta}^2)}{\eta^8},
\end{equation}

 Using
Eqs.(\ref{Chap}), (\ref{rhochap}) and (\ref{cc}) we can obtain
\begin{eqnarray}\label{khafan2}
w_D&=&\frac{-A}{\rho_D^{ch}}=\frac{-A}{A+Ba^{-6}}\\ \nonumber &&=
-1+\frac{2}{3}\frac{\eta(2D_{\eta}-3n^2M_p^2\eta^2-\alpha)\sqrt{3M_p^2\Omega_D}}{D_{\eta}^{3/2}}-\frac{b^2(1+\Omega_k)}{\Omega_D}.
\end{eqnarray}
Substituting Eq.(\ref{B2}) in (\ref{khafan2}), the constant $A$
can be obtain as
\begin{eqnarray}
A=\frac{1}{3a\eta^8}\label{A2}
\left[2\eta\sqrt{3M_p^2\Omega_DD_{\eta}}(-2D_{\eta}+3n^2M_p^2T^2+\alpha)+3aD_{\eta}^2(1+b^2\frac{1+\Omega_k}{\Omega_D})\right].
\end{eqnarray}
Substituting $A$ and $B$ in Eqs.(\ref{rhophi}) and (\ref{vphi}), we
can rewrite the scalar potential and kinetic energy terms for
interacting ECNADE model as follows
\begin{equation}\label{vphi2}
V(\phi)=\frac{\sqrt{\frac{3M_p^2\Omega_D}{D_{\eta}}}}{3a\eta^3}(-2D_{\eta}+3n^2M_p^2\eta^2+\alpha)+\frac{D_{\eta}}{2}(\frac{2+b^2\frac{1+\Omega_k}{\Omega_D}}{\eta^4}).
\end{equation}

\begin{equation}\label{dotphi2}
\dot{\phi}=\sqrt{\frac{2}{3a\eta^3}\sqrt{\frac{3M_p^2\Omega_D}{D_{\eta}}}(+2D_{\eta}-3n^2M_p^2\eta^2-\alpha)-\frac{b^2}{\eta^4}D_{\eta}\frac{1+\Omega_k}{\Omega_D}}.
\end{equation}
Using $\dot{\phi}=\phi^{\prime}H$, it is clear that
\begin{equation}\label{phi_prime1}
\phi^{\prime}=\frac{1}{H}\sqrt{\frac{2}{3a\eta^3}\sqrt{\frac{3M_p^2\Omega_D}{D_{\eta}}}(+2D_{\eta}-3n^2M_p^2\eta^2-\alpha)-\frac{b^2}{\eta^4}D_{\eta}\frac{1+\Omega_k}{\Omega_D}}.
\end{equation}
Integrating Eq.(\ref{phi_prime1}), we have the evolutionary behavior
of the scalar field as
\begin{equation}\label{eq_phi}
\phi(a)-\phi(a_0)=\int_{\ln{a_0}}^{\ln{a}}
\frac{1}{H}\sqrt{\frac{2}{3a\eta^3}\sqrt{\frac{3M_p^2\Omega_D}{D_{\eta}}}(+2D_{\eta}-3n^2M_p^2\eta^2-\alpha)-\frac{b^2}{\eta^4}D_{\eta}\frac{1+\Omega_k}{\Omega_D}}d\ln{a}.
\end{equation}
where $a_0$ is a present value of scale factor. Here we have
established the correspondence between the interacting ECNADE with
the Chaplygin gas dark energy and reconstruct the potential and the
dynamics of interacting ENCADE Chaplygin gas. It is also emphasized
that in the limiting case of $\alpha=\beta=0$ and $b=0$, the
relations (\ref{cc}, \ref{B2}, \ref{A2}, \ref{vphi2}, \ref{dotphi2},
\ref{phi_prime1} and \ref{eq_phi}) reduce to Eqs.(34, 36, 38, 40,
41, 42 and 43), respectively, in Ref.\cite{shyekh_age}.
\section{Interacting ECNADE generalized Chaplygin gas\label{Gen}}
In this section we extend this work to the interacting ECNADE
generalized Chaplygin gas .The equation of state of generalized
Chaplygin gas \cite{Ben}, is given by
\begin{eqnarray}\label{Chap2}
p_D^{gch}=\frac{-A}{{\rho_D^{\delta}}}.
 \end{eqnarray}
Substituting the above equation into the relativistic energy
conservation equation leads to a density evolving as
\begin{eqnarray}\label{rhochap2}
\rho_D^{gch}=\left({A+{B}a^{-3\gamma}}\right)^{1/\gamma},
\end{eqnarray}
where $\gamma=\delta+1$. Hence, it is easy to find the EoS parameter
of generalized Chaplygin gas as
\begin{eqnarray}\label{wchap2}
w_D^{gch}=\frac{p_D^{gch}}{\rho_D^{gch}}=\frac{-A}{{A+{B}a^{-3\gamma}}}.
\end{eqnarray}
Now we establish the connection between the interacting ECNADE model
and generalized Chaplygin gas dark energy. Equating Eq.(\ref{ss2})
with Eq.(\ref{rhochap2}) and also Eq.(\ref{wchap2}) with
Eq.(\ref{cc}), the constant parameters $A$ and $B$ can be obtained
as
\begin{equation}
B=[(\frac{D_{\eta}}{\eta^4})^{\gamma}-A]a^{3\gamma}.
\end{equation}

\begin{equation}
A=\frac{1}{3a}(\frac{\sqrt{D_{\eta}}}{\eta^4})^{\gamma}\left(2\eta\sqrt{\frac{3M_p^2\Omega_D}{D_{\eta}^{3/2}}}(-2D_{\eta}+3n^2M_p^2\eta^2+\alpha)+3aD_{\eta}(1+b\frac{1+\Omega_k}{\Omega_D})\right)
\end{equation}
Substituting $A$ and $B$ in Eqs.(\ref{rhophi}) and (\ref{vphi}), we
re-write the scalar potential and kinetic energy terms for
interacting ECNADE generalized Chaplygin gas as follows
\begin{eqnarray}\label{v_phi_gen}
V(\phi)&=&\frac{\sqrt{\frac{3M_p^2\Omega_D}{D_{\eta}^3}}}{3a}(\frac{D_{\eta}}{\eta^4})^{\gamma-\delta}\eta(-2D_{\eta}+3n^2M_p^2\eta^2+\alpha)\\ \nonumber&&+\frac{1}{2}(\frac{D_{\eta}}{\eta^4})^{\gamma-\delta}
(1+b^2\frac{1+\Omega_k}{\Omega_D})+\frac{1}{2}(\frac{D_{\eta}}{\eta^4})
\end{eqnarray}
\begin{eqnarray}\label{phi_dot_gen}
\dot{\phi}=\sqrt{\frac{2}{3a}(\frac{D_{\eta}}{\eta^4})^{\gamma-\delta}\sqrt{\frac{3M_p^2\Omega_D}{D_{\eta}^3}}\eta(2D_{\eta}-3n^2M_p^2\eta^2-\alpha)+\frac{D_{\eta}}{3\eta^4}-\frac{1}{3}(\frac{D_{\eta}}{\eta^4})^{\gamma-\delta}(1+b^2\frac{1+\Omega_k}{\Omega_D})}
\end{eqnarray}
Using $\dot{\phi}=\phi^{\prime}H$, we can find
\begin{eqnarray}\label{phi_prime_gen}
\phi^{\prime}&=& \\&&\frac{1}{H}
\nonumber\sqrt{\frac{2}{3a}(\frac{D_{\eta}}{\eta^4})^{\gamma-\delta}\sqrt{\frac{3M_p^2\Omega_D}{D_{\eta}^3}}\eta(2D_{\eta}-3n^2M_p^2\eta^2-\alpha)+\frac{D_{\eta}}{3\eta^4}-\frac{1}{3}(\frac{D_{\eta}}{\eta^4})^{\gamma-\delta}(1+b^2\frac{1+\Omega_k}{\Omega_D})}
\end{eqnarray}
 Integrating
Eq.(\ref{phi_prime_gen}), we obtain the evolutionary behavior of the
scalar field as
\begin{eqnarray}\label{gen_evol}
&\phi(a)&-\phi(a_0)= \\ &&
 \int_{\ln{a_0}}^{\ln{a}} \frac{1}{H}\nonumber\sqrt{\frac{2}{3a}(\frac{D_{\eta}}{\eta^4})^{\gamma-\delta}\sqrt{\frac{3M_p^2\Omega_D}{D_{\eta}^3}}\eta(2D_{\eta}-3n^2M_p^2\eta^2-\alpha)+\frac{D_{\eta}}{3\eta^4}-\frac{1}{3}(\frac{D_{\eta}}{\eta^4})^{\gamma-\delta}(1+b^2\frac{1+\Omega_k}{\Omega_D})}d\ln{a}
\end{eqnarray}
where $a_0$ is a present value of scale factor. Here we have
established the corespondence between the interacting ECNADE with generalized Chaplygin gas model and reconstruct
the potential and the dynamics of generalized Chaplygin cosmology. It is worth noting that, by
putting $\gamma=2, \delta=1$, all the above equations for interacting ECNADE generalized Chaplygin gas
reduce into those for interacting ECADE Chaplygin gas.

\section{Conclusions\label{CONC}}
Here we considered the interacting entropy-corrected agegraphic dark
energy models with CDM in a non-flat universe. We established a
correspondence between the interacting ECADE density with the
Chaplygin gas energy in a non-flat universe. We reconstructed
the potential and the dynamics  of the interacting entropy corrected
agegraphic scalar field  which describe the Chaplygin cosmology. We
also derived the potential and dynamics of the interacting
entropy-corrected new agegraphic model which describe the  Chaplygin
cosmology. Finally, we have extended this work into the interacting
entropy-corrected new agegraphic scalar field which can describe the
generalized Chaplygin cosmology.

\end{document}